\documentclass[12pt,preprint]{aastex}

\shorttitle{Hyades Mass-[Fe/H] Relation}
\shortauthors{Aaron Dotter and Brian Chaboyer}

\newcommand{\Ms}{\mathrm{M_{\odot}}}
\newcommand{\Me}{\mathrm{M_{\oplus}}}

\begin{document}

\title{Stellar Pollution and [Fe/H] in the Hyades}
\author{Aaron Dotter and Brian Chaboyer}
\affil{Department of Physics and Astronomy, Dartmouth College, 6127 Wilder Laboratory,
 Hanover, NH 03755}
\email{Brian.Chaboyer@Dartmouth.edu}

\begin{abstract}
The Hyades open cluster presents a unique laboratory for planet
formation and stellar pollution studies because all of the stars have
essentially the same age and were born from the same cloud of gas.
Furthermore, with an age of $\sim$650 Myr most of the intermediate and
low mass stars are on the main sequence.  Given these assumptions, the
accretion of metal rich material onto the surface of a star during and
shortly after the formation of planetary systems should be evident via
the enhanced metallicity of the star.  Building on previous work,
stellar evolution models which include the effects of stellar
pollution are applied to the Hyades.  The results of several Monte
Carlo simulations, in which the amount of accreted material is drawn
at random from a Gaussian distribution with standard deviation equal
to half the mean, are presented.  An effective temperature--[Fe/H]
relation is produced and compared to recent observations.  The
theoretical predictions presented in this letter will be useful in
future searches for evidence of stellar pollution due to planet
formation.  It is concluded that stellar pollution effects at the mean
level of $\geq$2 $\Me$ of iron are ruled out by the current
observational data \citep{psc}.
\end{abstract}

\keywords{open clusters and associations: individual(Hyades) ---
planetary systems: protoplanetary disks ---
stars: abundances ---
stars: evolution }

\section{Introduction}

The concept of stellar pollution, that as the protoplantery disk
around a star forms planets and settles into a stable configuration
the star may accrete some of the metal rich material from the disk,
has potentially observable implications for the parent star. The
magnitude of the observable implications can vary widely, being
influenced by the planetary formation process, which depends on the
composition and dynamics of the cloud from which the star is born, as
well as the depth of the surface mixed layer of the star, which in
turn depends on the mass of the star. \citet{mea} compared a sample of
$\sim$500 main sequence stars in the Solar neighborhood to stellar
evolution models. Their analysis showed that the observations are
consistent with the stars having accreted 0.5 $\Me$ of iron on
average. The sample consisted of field stars and thus the signal for
pollution had to be disentangled from variations in age and bulk
metallicity.

Based on the proposal by \citet{quillen} that scatter about the main
sequence in the Hyades color-magnitude diagram (CMD) of \citet{deB}
can be used to limit the level of pollution, \citet[hereafter DC]{dc}
applied stellar evolution models which incorporate the effects of
stellar pollution to the Hyades. The authors found that pollution on
the level of $\sim$1.5 $\Me$ could be ruled out by scatter in the
CMD. A more direct test of pollution in the Hyades, presented in this
letter, involves comparing polluted stellar evolution models to
observations of [Fe/H] in Hyades stars.  Such a test is possible with
the high precision [Fe/H] data set presented by \citet[hereafter
PSC]{psc}. At present, at least three other groups are known to be at
work on similar observations in the Hyades \citep{boe,ful,pea}. These
observations may have the precision and wide effective temperature
range to make a definitive statement about stellar pollution in the
Hyades. The work of PSC is, within the quoted uncertainties,
consistent with zero dispersion about the mean [Fe/H] of the
sample. However, as shown below, it is also in good agreement with the
predictions drawn from polluted stellar evolution models.

\section{Monte Carlo Simulations}

Four distinct Monte Carlo simulations were performed using Chaboyer's
stellar evolution program with modifications to deal with stellar
pollution. Each simulation uses the same set of 77 stellar models
described by DC. These stars lie in the effective temperature range of
3500 K to 8000 K which corresponds to masses in the range of roughly
0.5 to 2 $\Ms$. Each Monte Carlo simulation consists of 1000
individual runs through the set of polluted stellar models. For each
stellar model in a given run, the amount of polluting material is
drawn at random from a Gaussian distribution with standard deviation
equal to half the mean. \citet{mea} find that, on average, field stars
in the Solar neighborhood have accreted $\sim$0.5 $\Me$ of iron. Three
of the four simulations have means of 0.5, 1, and 2 $\Me$ of iron. The
fourth simulation begins with a mean of 0.5 $\Me$ but adds in a 5\%
probability that a star will have giant planet in a tight orbit
leading to an average accretion of 5 $\Me$ of iron \citep{mc}. In
order that the simulations have the mean [Fe/H] = 0.13 (within a few
hundredths of a dex) in agreement with observations \citep[PSC]{bf} the
models have the same initial bulk metallicity, Z=0.024. The only
exception is the 2 $\Me$ case for which all models have a lower
initial bulk metallicity (Z=0.020) in order to keep the mean [Fe/H]
reasonably close to the observed value.

The main goal of the simulations is to determine how [Fe/H] varies
with stellar mass and pollution. To obtain the mass--[Fe/H] relation
from one simulation, the mass--[Fe/H] relation is created for each of
the individual runs.  A smooth curve is fit to each individual
relation using the LOWESS (Locally Weighted Regression) technique
\citep{cleve}. Finally, all of the 1000 individual mass--[Fe/H] curves
are averaged.  The process is analogous to the CMD analysis of DC. Color
tables from \citet{kur} are used for the color transformations.  

To compute overall [Fe/H] statistics for each of the four simulations,
the mean [Fe/H] value is computed from the 77 stellar models in each
individual run. The overall mean is the average of all 1000 individual
means. The [Fe/H] statistics from each simulation are also divided
into four mass bins. Within each bin the mean and standard deviation
about the mean are determined from all of the stellar models which
have masses in the given range.

\section{Results}

\begin{deluxetable}{ccccccc}
\tabletypesize{\footnotesize}
\tablecolumns{7}
\tablewidth{0pc}
\tablecaption{Monte Carlo Results\label{tab1}}
\tablehead{
\colhead{}&\colhead{}&\colhead{}&\multicolumn{4}{c}{$\Delta$[Fe/H]}\\

\cline{4-7}\\
\colhead{}&\colhead{}&\colhead{}
&\colhead{$M_*<$0.6 $\Ms$}&\colhead{0.6--1.0 $\Ms$} 
&\colhead{1.0--1.4 $\Ms$}&\colhead{1.4--1.8 $\Ms$}\\

\colhead{}&\colhead{}&\colhead{} 
&\colhead{$\mathrm{T_{eff}}<$4000 K}&\colhead{4000--5400 K}&\colhead{5400--6500 K}&\colhead{6500--8000 K}\\

\colhead{No.}&\colhead{Accreted Mass ($\Me$ Iron)}&\colhead{Initial Z}
&\colhead{B--V$>$1.35} 
&\colhead{0.74--1.35}&\colhead{0.43--0.74}&\colhead{0.23--0.43}
}

\startdata

1&0.50$\pm$0.25&0.024&0.000$\pm$0.002&0.001$\pm$0.004&0.017$\pm$0.019&0.086$\pm$0.079\\
2&\#1 (95\%), 5.00$\pm$2.50 (5\%)&0.024&0.000$\pm$0.012&0.002$\pm$0.018&0.026$\pm$0.060&0.107$\pm$0.152\\
3&1.00$\pm$0.50&0.024&0.000$\pm$0.005&0.004$\pm$0.008&0.040$\pm$0.036&0.157$\pm$0.137\\
4&2.00$\pm$1.00&0.020&0.000$\pm$0.012&0.012$\pm$0.022&0.095$\pm$0.079&0.312$\pm$0.248

\enddata

\end{deluxetable}

The primary results of this paper are presented in Table \ref{tab1},
which presents theoretical predictions for the mean [Fe/H] and its
dispersion as a function of stellar mass.  These predictions are
presented as relative abundances rather than absolute because relative
abundances are immediately obtained from observational data whereas
absolute abundances introduce another source of uncertainty, namely
the Solar abundances. For this reason, among others, one study may
have a different mean value of [Fe/H] than another.  For the purposes
of testing for the existence of stellar pollution, it is the behavior
of the relative abundances that matters most.  At each location in the
$\Delta$[Fe/H] portion of Table \ref{tab1} the first value is the mean
[Fe/H] value for that range of models relative to the mean for the low
mass models. The second value (following the ``$\pm$'') is the
standard deviation about the mean as described at the end of $\S$2. In
a given row, the mean value represents the difference between one mass
range and another, and the standard deviation of the mean gives an
idea of the dispersion within one mass range. Overall absolute
abundances for the four simulations listed in Table \ref{tab1} are:
$<$[Fe/H]$>$=0.131$\pm$0.052 (1), 0.142$\pm$0.085 (2), 0.161$\pm$0.092
(3), and 0.155$\pm$0.176 (4).

These predictions reflect the differences in [Fe/H] exhibited by the
four cases outlined in the previous section and differences within
each case as a function of stellar mass. It is evident that assuming
5\% of the Hyades stars have giant planets in tight orbits (case 2)
increases $<$[Fe/H]$>$ by only 0.011 (8\% relative change) but
increases the standard deviation about the mean by 0.033 (63\%) over
the entire mass range covered in the simulations.  Thus a small
probability of having a giant planet in a tight orbit will have
relatively little influence on the average metallicity of a stellar
population but will have a significant influence on the scatter. The
[Fe/H] data is broken into mass bins in order simplify comparisons
with observational data. The Monte Carlo simulations indicate that, if
stellar pollution is the main factor in metallicity variations, then
stars above roughly 6500 K (1.4 $\Ms$) will have a significantly
higher mean [Fe/H] and dispersion than stars below 6500 K. The lower
the initial bulk metallicity the more pronounced this effect will be.

\begin{figure}
\plotone{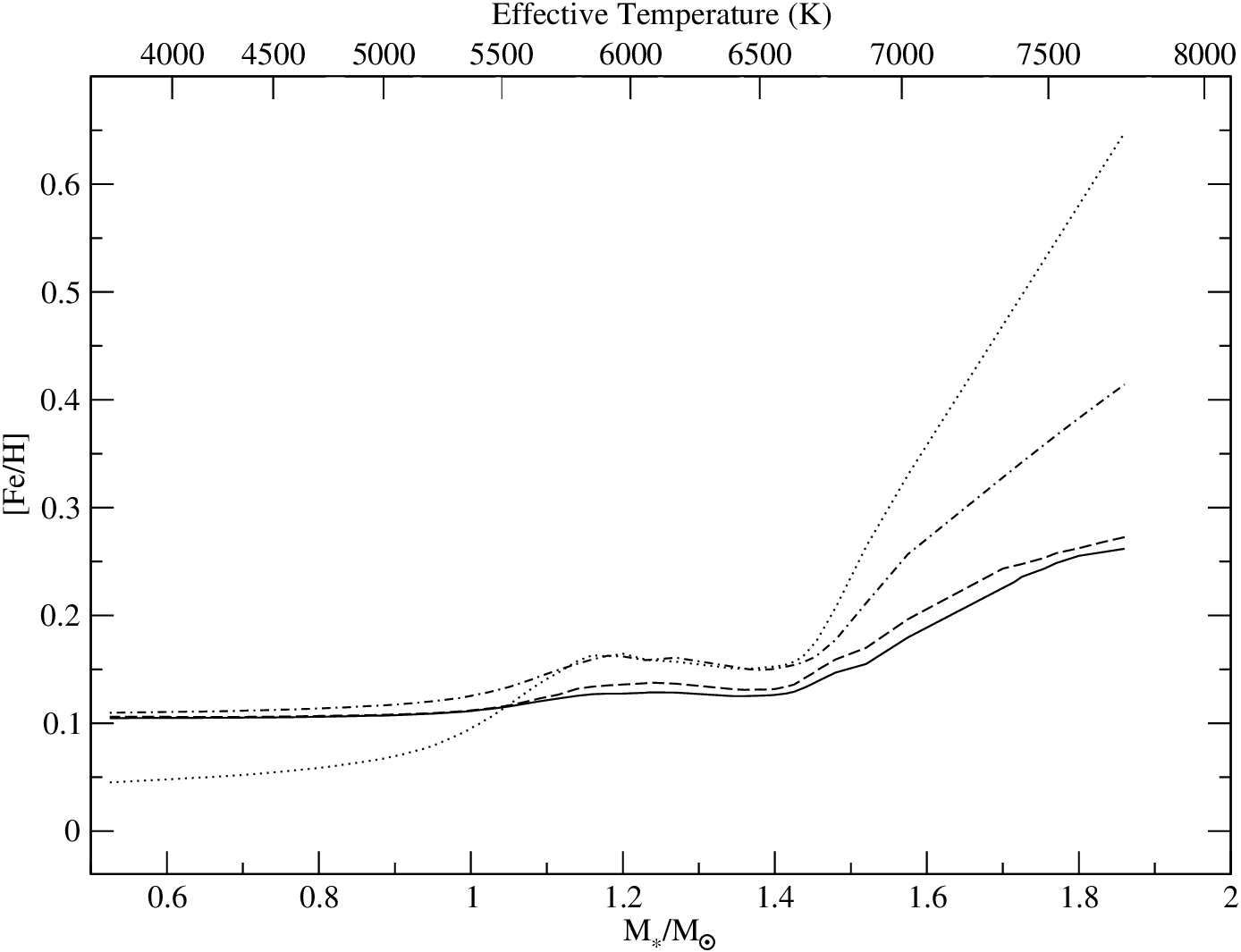}
\caption{The mass/$\mathrm{T}_{\mathrm{eff}}$--[Fe/H] relation for the
Hyades. The solid line is the 0.5 $\Me$ case (1 in Table
\ref{tab1}), the dashed line is the 0.5 $\Me$ with giant planets case
(2), the dash-dotted line is the 1.0 $\Me$ case (3), and the
dotted line is the 2.0 $\Me$ case (4).
\label{mfeh}}
\end{figure}

The mean mass--[Fe/H] relation is displayed for all four simulations
in Figure \ref{mfeh}.  The general trend is that [Fe/H] increases with
stellar mass which is consistent with the decreasing mass of the
surface mixed layer. The decrease at around 1.4 $\Ms$ is due to an
artificially increased surface mixed layer, for details see
\citet{mea}.

\section{Comparison with Observational Data}

The hotter temperatures and higher rotational velocities of A and F
stars make the task of measuring metal abundances exceedingly
difficult. As such, a detailed comparison of observations of A and F
stars to the Monte Carlo simulations is not currently
possible. However, comments are made regarding two studies of F and A
stars in the Hyades. \citet{bf} list effective temperatures and [Fe/H]
values for 14 F stars in the Hyades. The uncertainties in [Fe/H] are
all of order 0.10 dex and the authors find no [Fe/H] trend with
temperature. Nevertheless, a simple comparison suggests that, within
the uncertainties, the polluted models are in agreement with their
data.

\citet{vm} present a study of 29 F and 19 A stars in the Hyades. The 
authors provide uncertainties in their [Fe/H] values for only about 10
of the 48 stars, thus an analysis of the dispersion is not possible.
Furthermore, the data varies widely over a range of $\sim$0.6 dex in
[Fe/H] so simple averages in several temperature ranges like those in
Table \ref{tab1} would favor the outlying points. To avoid giving
equal weight to the outliers in the average values, the averages are
taken from the LOWESS fit to the data. (LOWESS weights each point
based on how close it is to the local average.)  In keeping with the
type of analysis performed by DC and in the following paragraphs, the
absolute values of the relative [Fe/H] values (relative meaning the
overall average has been subtracted off) are used. The absolute value
of the $\Delta$[Fe/H] data of \citet{vm} has a LOWESS-average value of
0.05 dex between 6000 and 6500 K, and 0.12 dex between 6500 and 8000
K. Comparing these data to Table \ref{tab1}, LOWESS-average values in
between 6000 and 6500 K are 0.04 dex (1), 0.06 dex (2), and 0.10 dex
(3). Between 6500 and 8000 K the simulations have LOWESS-average
values of 0.06 dex (1), 0.07 dex (2), and 0.17 dex (3). At the 90\%
confidence level, the \citet{vm} data is consistent with stellar
pollution at and below the 1 $\Me$ level but is inconsistent with
pollution at the 2 $\Me$ level.

PSC perform spectroscopic abundance analyses of 55 Hyades stars which
range in effective temperature from about 5000 to 6500 K. These
authors list relative [Fe/H] data and perform detailed error analysis
making it possible to directly compare the data to the predictions of
the polluted stellar models.  The star chosen as the standard about
which the relative abundances are determined (HD 35768) has an [Fe/H]
value somewhat below the cluster average so that the $\Delta$[Fe/H]
values average to $\sim$0.03 dex. For the purposes of this letter,
0.03 dex has been subtracted from each $\Delta$[Fe/H] value so that
$<$$\Delta$[Fe/H]$>$=0.  The modified $\Delta$[Fe/H] data of PSC can
be compared to the Monte Carlo simulations by constructing
color--[Fe/H] curves as described above.  The absolute value of the
$\Delta$[Fe/H] data is fit by the LOWESS technique for both the
observational data and the four simulations. Relative abundances for
the simulations are obtained by subtracting from each data point the
average [Fe/H] of the run in the same temperature range as the
observational data which corresponds to stellar masses between 0.8 and
1.25 $\Ms$.  PSC report an average error about the mean of 0.04 dex in
$\Delta$[Fe/H].

\begin{figure}
\plotone{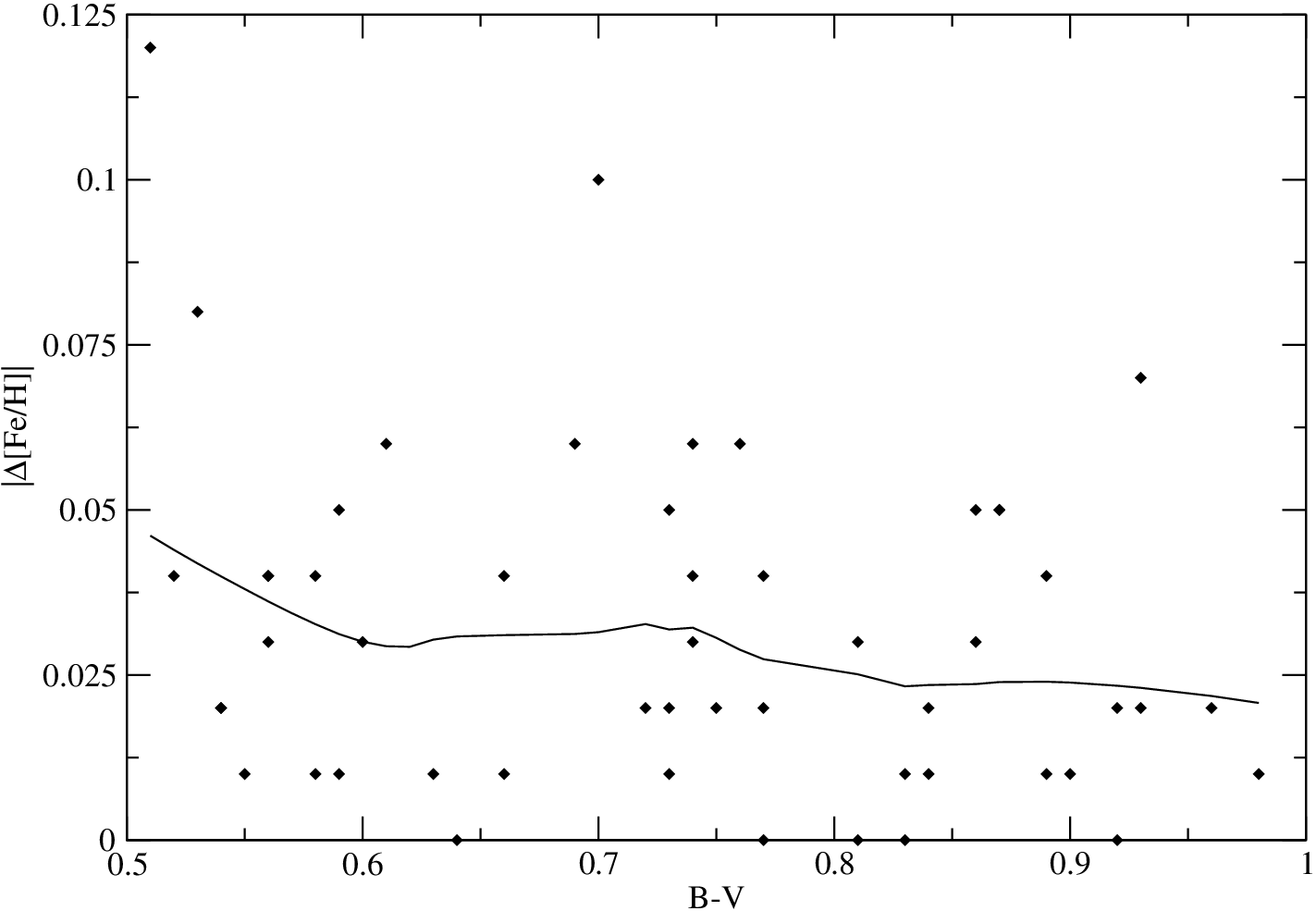}
\caption{
The absolute values of the $\Delta$[Fe/H] data of PSC and the LOWESS
fit. Three outlying points have been omitted from the plot for scaling
purposes. See the text for an explanation.
\label{psc1}}
\end{figure} 

\begin{figure}
\plotone{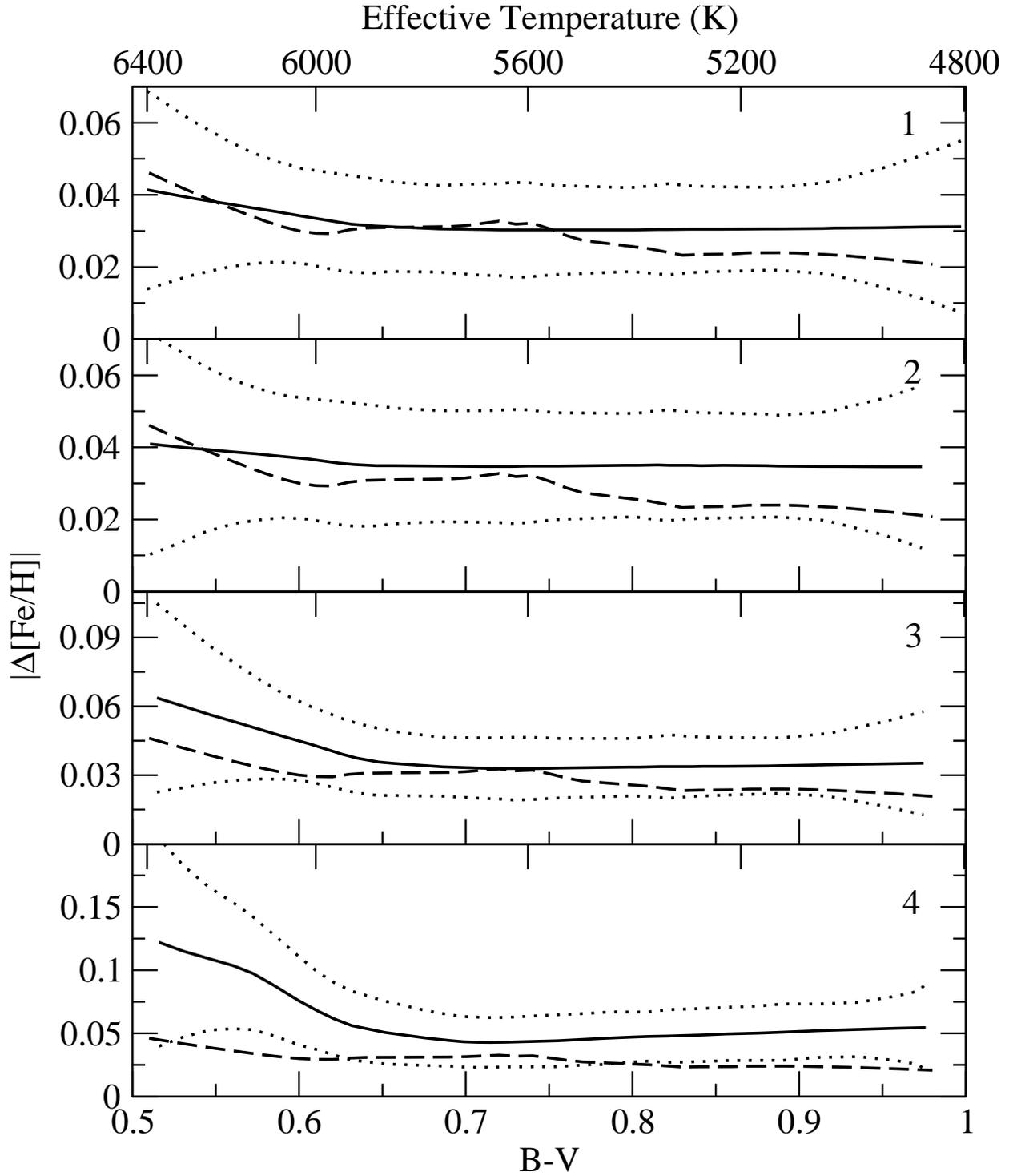}
\caption{Trends in $|\Delta$[Fe/H]$|$ from the Monte Carlo simulations
and the PSC data. Lines in each plot represent: the LOWESS fit to the
PSC data (dashed line), the Monte Carlo mean (solid line), and Monte
Carlo 90\% confidence levels (dotted lines). The number in the upper
right of each panel refers to Table \ref{tab1}.
\label{psc2}}
\end{figure}

Figure \ref{psc1} shows the absolute values of the modified
$\Delta$[Fe/H] data of PSC along with the LOWESS fit. The locations of
three stars have been omitted.  The stars are HD 20430, HD 20439, and
HD 14127 with (modified) $|\Delta$[Fe/H]$|$ = 0.18, 0.17, and 0.23,
respectively. That these stars are indeed members of the Hyades is
questionable, see the discussion in PSC.  In any case, because of
their large deviations these three stars play an insignificant role in
the LOWESS fit. It must be noted here that to within the quoted
average uncertainty of 0.04 dex the LOWESS fit is consistent with zero
scatter in [Fe/H].

A comparison of the results of the four Monte Carlo simulations to the
observational data is presented in Figure \ref{psc2}. In order to
directly compare the Monte Carlo simulations with the observational
data an additional step was added. The simulations represent intrinsic
values untouched by uncertainties involved in observations.  To
account for the observational uncertainties, the [Fe/H] value of each
star was drawn from a Gaussian distribution with mean given by the
simulation and a 0.04 dex standard deviation (in accordance with the
average error bar of PSC).  The individual graphs are numbered in the
upper right hand corners, the numbers refer to Table
\ref{tab1}. Stellar pollution at the 0.5 $\Me$ level cannot be ruled
out by the observational data, with (2) or without (1) the possibility
of having a giant planet in a tight orbit. Likewise with the 1 $\Me$
case (3). The 2 $\Me$ case (4), however, is not consistent with the
data at the 90\% confidence level. Thus the PSC data set rules out
pollution at the 2 $\Me$ level. The divergence of the 90\% confidence
levels at the high and low end of each panel is due to the exclusion
of the data points outside the limits considered. This behavior is
spurious and should not be considered part of the prediction, however,
for the purpose of comparing the simulations to the observational data
it is necessary to leave it as is. The B--V color is used as the
abscissa in the comparison presented here but effective temperature
could also have been used, the basic trends and conclusions are the
same.

The data of PSC cover stars between roughly 0.8 and 1.25 $\Ms$ but, as
Figure \ref{mfeh} and Table \ref{tab1} indicate, these stars do not
exhibit the greatest dispersion due to pollution effects.  A careful
analysis of hotter stars will be necessary to determine the level of
stellar pollution in the Hyades.

\section{Conclusions}
The formation of planets around a star may lead to the accreation of
metal rich material by the star.  Due to variations in the planet
formation process and variations of the mass of the surface mixed
layer with stellar mass, the effect of such stellar pollution will
vary from star to star. In this paper, the effects of stellar
pollution on the Hyades is studied.  The main result of this study is
presented in Table \ref{tab1} where the predicted mean [Fe/H] and
dispersion is given as a function of stellar mass. The predictions
have been binned by stellar mass in order to be more easily comparable
to observational data sets.  As stellar mass increases, increases in
both the mean [Fe/H] and the dispersion caused by stellar pollution
should be evident. A small probability that a star will have a giant
planet in a tight orbit has little impact on the average [Fe/H] but
does have a noticeable effect on the dispersion. An observational
program that can provide conclusive evidence for or against stellar
pollution in the Hyades needs precise abundances over a wide range of
spectral types. A significant increase in the mean [Fe/H] should arise
as the effective temperature increases from below to above 6500 K. In
addition, the dispersion is larger at higher temperatures. It will be
important to have relative abundances for both cool and hot stars so
that a trend can be drawn over a broad range of effective
temperatures. The present data sets rule out pollution at the $\ga 2
\Me$ level but are consistent with pollution at and below the 1 $\Me$
level.

\acknowledgments

\end{document}